\documentclass{iopart}
\usepackage[english]{babel}
\usepackage[utf8x]{inputenc}
\usepackage[T1]{fontenc}
\usepackage{amssymb}
\usepackage{graphicx}

\newtheorem{theo}{Theorem}

\newenvironment{my_enumerate}
{\begin{enumerate}
  \setlength{\itemsep}{1pt}
  \setlength{\parskip}{0pt}
  \setlength{\parsep}{0pt}}
{\end{enumerate}}

\newcommand{\bm}{\mathbf}
\newcommand{\stack}[2]{\begin{array}{c}{#1}\\{#2}\end{array}}

\begin{document}
\title{Constructing and sampling directed graphs with given degree sequences}
\author{H Kim$^{1,2}$, C~I Del~Genio$^3$, K~E Bassler$^{4,5}$ and Z Toroczkai$^2$}
\address{$^1$ Department of Physics, Virginia Tech, Blacksburg, VA, 24061, USA}
\address{$^2$ Interdisciplinary Center for Network Science and Applications (iCeNSA),
Department of Physics, University of Notre Dame, Notre Dame, IN 46556, USA}
\address{$^3$ Max-Planck-Institut f\"ur Physik Komplexer Systeme, N\"{o}thnitzer 
Str.~38, D-01187 Dresden, Deutschland}
\address{$^4$ Department of Physics, University of Houston, 617 Science and 
Research 1, Houston, Texas 77204-5005, USA}
\address{$^5$ Texas Center for Superconductivity, University of Houston, 202 
Houston Science Center, Houston, Texas 77204-5002, USA}
\ead{toro@nd.edu}
\begin{abstract}
The interactions between the components of complex networks are often
directed. Proper modeling of such systems frequently requires the construction 
of ensembles
of digraphs with a given sequence of in- and out-degrees.
As the number of simple labeled graphs with a
given degree sequence is typically very large even for short
sequences, sampling methods are needed for statistical studies. 
Currently, there are two main classes of methods that generate samples.
One of the existing methods
first generates a restricted class of graphs, then uses a
Markov Chain Monte-Carlo algorithm based on edge swaps to
generate other realizations. As the mixing time of this process
is still unknown, the independence of the samples is not well 
controlled. The other class of methods is based on the 
Configuration Model that may lead to unacceptably many 
sample rejections due to self-loops and multiple edges. Here we 
present an algorithm that can directly construct all possible realizations 
of a given bi-degree sequence by
simple digraphs. Our method is rejection free, guarantees the independence of the
constructed samples, and provides their weight. The weights can then
be used to compute statistical averages of network observables as
if they were obtained from uniformly distributed sampling, or from any other
chosen distribution.
\end{abstract}
\maketitle 

\section{Introduction and definitions}
In network modeling problems~\cite{New10,Eas10,Bar08,New06,Boc06,Ben04,Dor02},
one often needs to generate ensembles of graphs obeying a given constraint. A typical 
constraint is the case when the only information available is the degrees of the nodes, 
and not the actual connectivity matrix. 
Note that the node degrees by themselves, that is
the {\em degree sequence}
in general does not determine a graph uniquely: there can be a very large number
of graphs having the same degree sequence~\cite{Ben78}. Full graph connectivity is 
uniquely determined by the degree sequence only for a special class of sequences
(see Ref.~\cite{Kor76} for the case of undirected graphs).  

Often, the interest lies in the study of network observables, \emph{as determined}
by the given sequence of degrees, and unbiased by anything else. These can be
graph theoretical measures, or properties of processes happening on the network
(e.g., spreading processes, such as of opinion or disease). The problem of creating
and sampling graphs with a given degree sequence, i.e., \emph{degree-based graph
construction}~\cite{Kim09,Del10}, is a well-known and challenging problem that
has attracted considerable interest amongst researchers 
\cite{Ben78,Kim09,Del10,Bol80,Tay82,Mol95,Rao96,Kan99,New01,
Chu02,Mas02,Mil02,Itz03,Mil03,Par03,Vig05,Bri06,Coo07,
Bia08,Bia09,Erd09}. There are two main classes of algorithms that are used today to
achieve the construction of graphs with given degree sequences. One of them is
typically referred to as ``switching" or edge-swap based~\cite{Tay82,Rao96,Kan99,Vig05,Coo07},
while the other one is usually called ``matching'' or stub-matching based~\cite{Ben78,Bol80,Mol95,New01,Bog04,Cat05,Ser05,Bri06,Blitz}.
Switching methods repeatedly swap the ends of two randomly chosen edges within
a Markov Chain Monte-Carlo (MCMC) scheme until a new, quasi-independent, sample
is produced. Unfortunately, the mixing time of MCMC schemes for arbitrary sequences
is not known in the general case. The other class consists of direct
construction methods, which perform pairwise matchings of the half-edges emanating
from randomly chosen nodes until all edges are realized. Unfortunately, this method
can easily generate multiple edges and self-loops, i.e., edges starting and ending
on the same node, after which the sample must be rejected in order to avoid 
biases~\cite{Har99}. For a comparison of the two classes of methods see 
Ref.~\cite{Mil03}.

Recently, a novel degree-based construction~\cite{Kim09} and sampling
method~\cite{Del10} was introduced for undirected graphs, which has a
worst-case scaling of  ${\cal O}(NM)$, where $M$ is the number of 
edges ($2M$ is the sum of the degrees, which are given). 
A similar method was obtained independently in Ref. \cite{Blitz}, but that 
method is less efficient, with a worst-case scaling of ${\cal O}(N^2M)$.
Although the algorithm in Ref.~\cite{Del10} 
is a direct construction method using stub-matchings, it is rejection
free,  the samples are statistically independent and the algorithm also
provides a weight for every realization.

In many systems the interaction between two entities
is not mutual but has a direction from one to the other, such as in
the cases of human relationships in social networks \cite{socnet}, gene interactions
in regulatory networks, trophic interactions in food webs \cite{Mil02,Mor03}, etc. 
Such systems 
require a representation by directed graphs (digraphs). In fact, undirected
graphs can be interpreted as digraphs in which there are two, oppositely
directed edges for each connected pair of nodes. Here we present a
generalization of the degree-based graph construction problem to directed
graphs. Some of the necessary mathematical foundations, laid down in
Ref.~\cite{Erd09}, are here used and expanded to introduce a digraph
construction and sampling algorithm. Although the approach follows
closely the one introduced by us for the undirected case~\cite{Del10},
the generalization is not at all straightforward, and there are significant
differences that the directed nature of the links induces.

\begin{figure}
\centering
\includegraphics[width=0.7\textwidth]{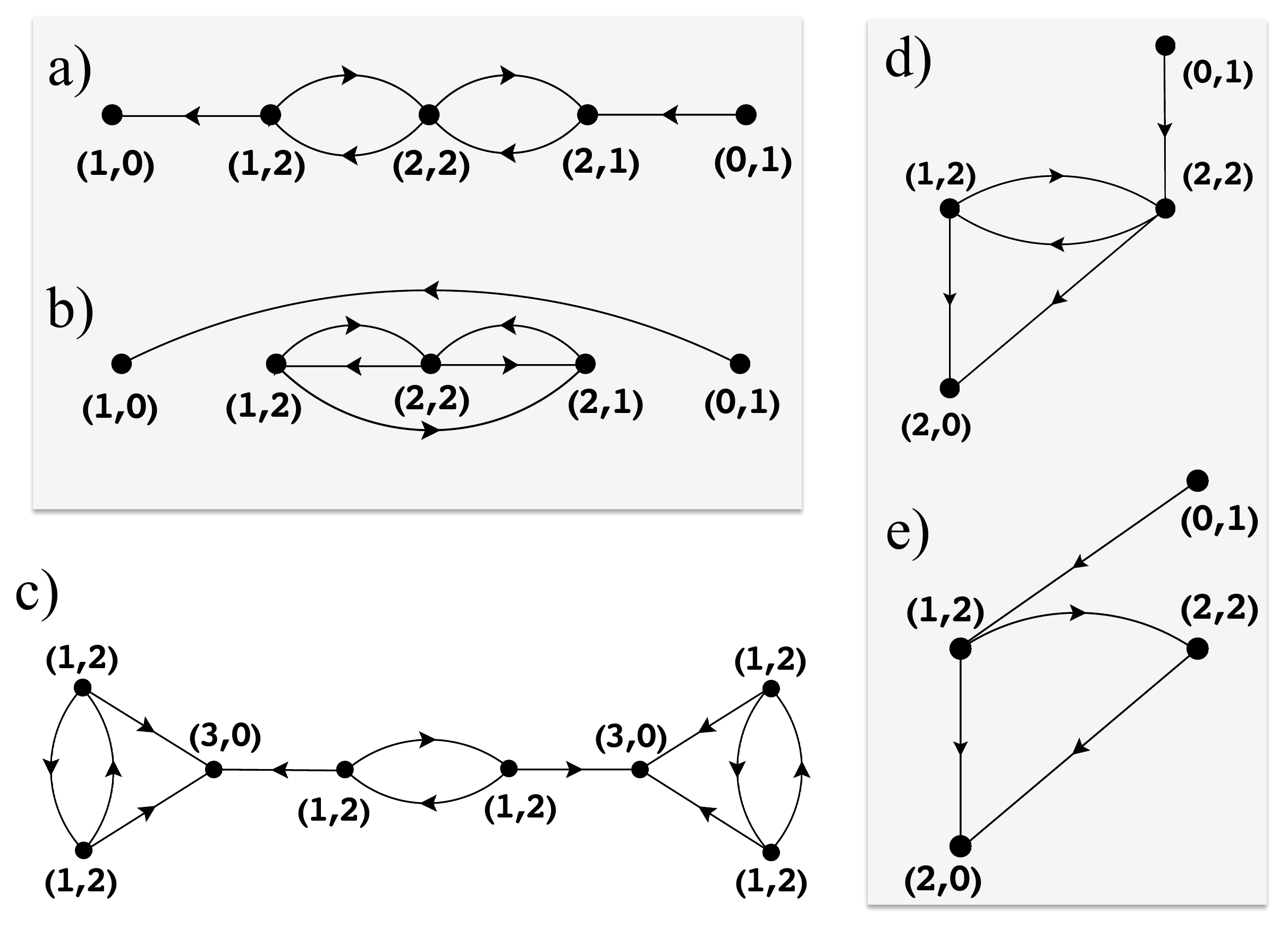}
\caption{Examples of realizations of graphical bi-degree sequences.
Panels a) and b) show two non-isomorphic realizations of the same bds.
Panel c) shows a digraph that cannot be obtained via the Havel-Hakimi
algorithm for digraphs. Panel d) shows a realization of a different bds.
Panel e) illustrates that not all possible connections lead to a simple
digraph even if a bds is graphical: in fact,  the connections in the
figure break the graphical character.}\label{fig:bds1}
\end{figure}

Before we present our algorithm, we introduce some notations, based on
Ref~\cite{Erd09}. Let us denote by $d^{(i)}_i$ and $d^{(o)}_i$ the in-\ and
out-degrees of a node $i$. Given the sequence
$\bm{D} = \left\lbrace\left(d^{\left(i\right)}_1, d^{\left(o\right)}_1\right), 
\left(d^{\left(i\right)}_2, d^{\left(o\right)}_2\right), \ldots, \left(d^{\left(i\right)}_{N}, 
d^{\left(o\right)}_{N}\right) \right\rbrace$
of non-negative integer pairs, we want to construct a \emph{simple} directed
graph $\bm{G}(V,\bm{E})$ such that node $k \in V$ has $(d^{(i)}_k, d^{(o)}_k)$
for its in- and out-degrees, respectively, for all $k=1,2,\ldots,N$.  A simple directed
graph is a graph that has no self-loops, nor multiple directed
edges in the same direction between two nodes. There can be at most two
edges between a pair of nodes, oppositely directed. We call the sequence
$\bm{D}$ a bi-degree sequence (bds for short). When there is a simple digraph
with a given bds $\bm{D}$ for its degrees, we say that the bds is \emph{graphical}
and that the digraph realizes $\bm{D}$.  Equivalently, we will also talk
about ``graphicality'' as a property. We distinguish realizations as \emph{labeled}
digraphs, and do not deal here with isomorphism questions. That is,
if two realizations are identical up to a permutation of their indices,
i.e., they are isomorphic, we will still consider them distinctly.
In order to avoid isolated nodes, in the following we will
assume that $d^{(i)}_j + d^{(o)}_j  > 0$, for all $j=1,\ldots,N$. As examples,
Figs.~\ref{fig:bds1}a) and~\ref{fig:bds1}b) show two realizations of the bds
$\bm{D}_1=\left\lbrace(1,0), (1,2), (2,2), (2,1), (0,1)\right\rbrace$,
and Fig.~\ref{fig:bds1}c) shows a realization of
$\bm{D}_2=\left\lbrace(3,0), (3,0), (1,2), (1,2), (1,2), (1,2), (1,2), (1,2)\right\rbrace$.
Examples of non-graphical bds are the sequences
$\bm{D}_3=\left\lbrace(2,2), (2,1), (1,3), (1,1)\right\rbrace$ and
$\bm{D}_4=\left\lbrace(5,6), (5,6), (5,6), (4,3), (3,3), (2,1), (2,1), (1,1)\right\rbrace$.

Notice that even if a bds is graphical, not all connection sequences
are guaranteed to end up with a simple digraph. For example, Fig.~\ref{fig:bds1}d)
shows a simple digraph realization of $\bm{D}_5=\left\lbrace(0,1), (2,0), (1,2), (2,2)\right\rbrace$.
However, if we were to place the first four edges as in Fig.~\ref{fig:bds1}e),
we would break graphicality: from there on, we would not be able to
complete the realization of the bds without creating either self-loops
or multiple edges. Hence, it is important to find an algorithm that builds digraphs
with a given bds. As we will see, this is a challenging problem in itself.

\begin{figure}
\centering
\includegraphics[width=0.7\textwidth]{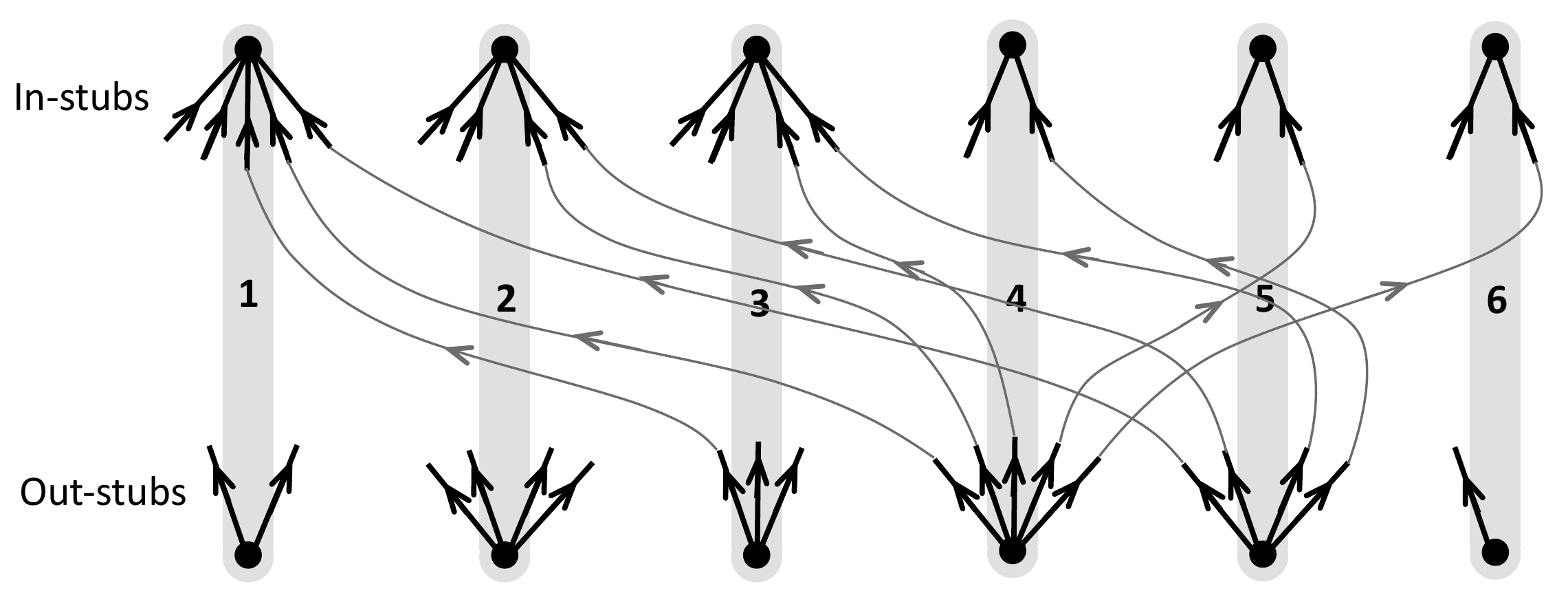}
\caption{The construction of a digraph realizing a given bds proceeds
by connecting the out-stubs of the nodes to the in-stubs of other nodes.
In this ``bipartite" representation the vertical grey bars represents single
nodes.}\label{fig:stubs}
\end{figure}

An algorithm that builds a digraph from a given bds sequentially connects
the out-links of a node to the in-links of others. We can think of these
out- and in-links as ``out-stubs'' and ``in-stubs'' emanating from a node,
that are paired up with the corresponding stubs of other nodes. An intuitive
representation of this is shown in Fig.~\ref{fig:stubs}.
As the graph construction algorithm proceeds, the number of stubs of the nodes decreases.
At any time during this process we will call the number of remaining in-stubs and out-stubs
of a node its \emph{residual} in- and out-degrees, and the corresponding bi-degree sequence
$\overline{\bm{D}} = \left\lbrace \left(\bar{d}^{\left(i\right)}_1, \bar{d}^{\left(o\right)}_1\right),
\left(\bar{d}^{\left(i\right)}_2, \bar{d}^{\left(o\right)}_2\right) \dots ,
\left(\bar{d}^{\left(i\right)}_{N}, \bar{d}^{\left(o\right)}_{N}\right) \right\rbrace$
the \emph{residual bds}.

Finally, another concept we will need to use in what follows is the notion of
\emph{normal order}~\cite{Erd09}, which is essentially the lexicographic order
on the bds. That is, we say that a bds is in normal order, if for all $1 \leq j \leq N-1$,
we have either $d^{(i)}_j > d^{(i)}_{j+1}$ or, if $d^{(i)}_j = d^{(i)}_{j+1}$,
then $d^{(o)}_j \geq d^{(o)}_{j+1}$. Thus, the bds
$\bm{D}_6 = \left\lbrace(5,2), (4,4), (4,3), (2,5), (2,4), (2,1)\right\rbrace$
shown in Fig.~\ref{fig:stubs}, is arranged in normal order. Once a bds is in normal
order, we will use the words `left' or `right' to describe the directions towards
lower or higher index values in the sequence.

The remainder of this paper is organized as follows: Section~\ref{mathf} introduces the fundamental mathematical
notions and algorithmic considerations that are at the basis of our digraph construction algorithm.
Section~\ref{thealg} presents the algorithm and its derivation details. Readers interested
only in the algorithm itself may skip Subsection~\ref{derivations} and proceed to the
summary described in the beginning of Section~\ref{thealg} and in Subsection~\ref{summary}.
Section~\ref{samprob} deals in detail with the digraph sampling problem, provides
the derivation of the sample weights and presents a simple example. Section~\ref{algcomp}
is dedicated to the complexity of the algorithm, and Section~\ref{discussion}
concludes the paper. 

\section{Mathematical foundations}\label{mathf}

As seen from the examples above, not all sequences of non-negative integer
pairs can be realized by simple digraphs. The sufficient and necessary conditions
for the realizability of a bds are given by the "FR" theorem~\cite{Lesniak86,Ful60,R63}:
\begin{theo}[Fulkerson-Ryser]\label{FR}
A sequence of non-negative integer pairs
$\bm{D}=\left\lbrace\left(d^{\left(i\right)}_1, d^{\left(o\right)}_1\right), \ldots, 
\left(d^{\left(i\right)}_N, d^{\left(o\right)}_N\right)\right\rbrace$
with $d^{(i)}_1 \geq d^{(i)}_2 \geq \ldots \geq d^{(i)}_N$
is graphical iff
\begin{eqnarray}
&&\!\!d^{(i)}_i \le N-1\;, \;\; d^{(o)}_i \le N-1, \;\; 1\leq i \leq N \label{ladegs}\\
&&\!\!\sum_{i=1}^{N} d^{(i)}_i = \sum_{i=1}^{N} d^{(o)}_i\;, \label{stubeq} \\
&&\!\!\!\!\!\!\!\!\!\mbox{and for all}\;\;1\leq k \leq N-1: \nonumber \\
&&\!\!\sum_{i=1}^{k} d^{(i)}_i\le \sum_{i=1}^{k}\min\left\lbrace k-1, d^{(o)}_i\right\rbrace  
\!+\!\!\!\! \sum_{i=k+1}^{N}\!\!\!\min\left\lbrace k, d^{(o)}_i\right\rbrace\;.\;\;\;\;\label{FReq}
\end{eqnarray}
\end{theo}
Given a bds, we can easily test if it is graphical using this theorem,
and thus we will also refer to it as the "FR test". Condition~(\ref{ladegs})
states that both the number of in- and out-degrees for all nodes must
be no larger than the number of other nodes it could connect to, or receive
connections from. Condition~(\ref{stubeq}) is a consequence of the requirement
that every out-stub must join an in-stub somewhere else; the sequence
$\bm{D}_3$ given in one of the above examples is not graphical because
it fails this condition. Condition~(\ref{FReq}) is less intuitive. Its
left hand side is the total number of in-stubs that the group of $k$ highest
in-degree nodes can receive. {\em Within this group}, a node's out-stubs can
absorb no more of those in-stubs from the same group than its out-degree or $k-1$ (it cannot
absorb from itself), whichever is smaller (giving the first sum on the rhs of (\ref{FReq})). 
{\em Outside of this group}, a node
cannot absorb more of those in-stubs than its out-degree or $k$, whichever is smaller
(the second sum on the rhs of (\ref{FReq})).
Hence, the necessity of~(\ref{FReq}). For the complete proof see Refs.~\cite{Ful60,R63}. Note that
the example sequence $\bm{D}_4$ above fails condition~\ref{FReq} for
$k=3$. The FR test is the directed version of the Erdős-Gallai (EG)
theorem (test) for undirected graphs.

An important note is that bi-degree sequences are \emph{less constraining}
than undirected ones. The out-stub of a node is always connected to an in-stub
of another, not affecting that node's out-stubs, whereas such distinction
does not exist for the undirected case. Alternatively, if we disregard for
a moment the directionality of the links and consider the degree of the node
to be the sum of its in- and out-degrees, then the corresponding graph realizing
the bds can have two edges running between the same pair of nodes, whereas
this is not allowed in the undirected case.

\subsection{Algorithmic considerations}
The FR theorem only tests for graphicality, but it does not provide an algorithm
for constructing the digraph(s) realizing the given bds. At first sight this might
not seem an issue. However, the sequence $\bm{D}_5$ in Figs.~\ref{fig:bds1}d 
and~\ref{fig:bds1}e)
reminds us that graphicality can easily be broken by a careless connection of stubs.
Clearly, for the purposes of digraph construction, it should not matter which
edges we create first, as long as we make sure that every connection made does not
break graphicality. In other words, the possibility to create the rest of the edges,
so that  a simple digraph results in the end, must always be preserved. Thus, the
key for the creation of an algorithm that builds simple digraphs realizing a given
bds without rejections is in a theorem that allows us to check if we would break
graphicality by placing a specific connection. Indeed, such theorems exist, and they
will be discussed below. However, interestingly, they require that connections be
made from the \emph{same node}, until all its stubs are used away into edges. That
is, assuming that we already made some connections from a given node $i$, preserving
graphicality, these theorems give necessary and sufficient conditions for keeping
graphicality by the next connection \emph{still} involving node $i$. Simply put,
they won't work in general, if we would attempt a new connection from $j$ to $k$, where $j,k\neq i$,
while node $i$ still has dangling stubs.

The connections already made from $i$ to some set of nodes ${\cal X}_i$ represent
a \emph{constraint} for the new connections from $i$, as these novel connections
must avoid the set ${\cal X}_i$. We call such a constraint associated to
a node a \emph{star constraint} on that node. Once all the stubs of node
$i$ are connected into edges while preserving graphicality, we obtain a graphical
residual sequence $\bm{D'}$ on at most $N-1$ nodes. Clearly, the new connections
we  make from this point on will not be constrained in any way by the connections
we made from node $i$. For the purposes of realizing the sequence $\bm{D'}$
we can just simply remove node $i$ with its fully completed connections,
create a realization by a simple graph of $\bm{D'}$, then, in the end, add
back node $i$ with its connections to this graph in order to obtain a realization
of $\bm{D}$. The comments above hold both for the undirected and directed cases.

One might think of using the EG test for the undirected case and the FR test
for the directed case on a residual degree sequence to decide if graphicality
was broken after attempting a new connection from the same node. For the \emph{undirected case},
we have shown in Ref.~\cite{Del10} that the passing of the EG test by the residual
sequence is only a \emph{necessary condition}, if there is already a star constraint
on a node. For example, consider the graphical degree sequence $\bm{d}=\left\lbrace6,5,5,3,3,2,1,1\right\rbrace$,
and assume that we made connections from node $i=3$ to nodes ${\cal X}_3 
= \left\lbrace1,6,7\right\rbrace$.
The residual sequence after these connections is 
$\bm{d'}=\left\lbrace5,5,2,3,3,1,0,1\right\rbrace$.
It is easy to check that it passes the EG test. However, we will break graphicality
with \emph{every} realization of $\bm{d'}$, because it will form a double edge with
one of the existing connections from node $i=3$ to ${\cal X}_3$. Thus, additional 
considerations have to be made to ensure the graphicality of the residual sequence 
for the undirected case, as described in \cite{Del10}. For the directed case
here we use the sufficient and necessary conditions for graphicality under star constraints
as provided by Theorem~\ref{star} below, proven in Ref.~\cite{Erd09}.

From now on, we will always talk about algorithms that first finish all the
out-stubs of a node before moving onto another node with non-zero out-degree.
In the case of a graphical bds, once all the out-degrees of all the nodes have
been connected into directed edges, we are guaranteed to have completed a digraph,
because the total number of in-stubs equals the total number of out-stubs,
according to property~(\ref{stubeq}).

\subsection{Theorems on which the algorithm is based}
An algorithm that builds graphical realizations of degree sequences of simple
\emph{undirected} graphs is the Havel-Hakimi (HH) algorithm~\cite{Hav55,Hak62}: we
choose any node with non-zero residual degree, then we connect all its stubs
to nodes with the largest residual degrees avoiding self and multiple connections.
This process is repeated with other nodes until all stubs of all nodes are used.
There is a corresponding version of the HH algorithm for bi-degree sequences
as well, introduced first in Ref.~\cite{Kle73}, then rediscovered independently
in Ref.~\cite{Erd09}, the latter providing an alternative proof. The HH algorithm
for bds proceeds as follows: given a normal-ordered bds, choose any node with 
non-zero residual out-degree, then connect all its out-stubs to nodes with the 
largest residual in-degrees, without creating multiple edges running in the same
direction, nor self-loops.
Reorder in normal order 
the residual sequence and repeat this process until all stubs of all nodes are used. While for any given
bds, the HH algorithm will construct a set of digraphs, it cannot construct
\emph{all possible} digraphs realizing the same sequence, as shown in Ref.~\cite{Erd09}.
For example, the HH algorithm can never result in the digraph shown in Fig.~\ref{fig:bds1}c)
realizing the example sequence $\bm{D}_2$ above. It  is easy to see why: there are
two kinds of nodes in this example, with bi-degrees $(3,0)$ and $(1,2)$. The only
nodes with non-zero out-degrees are the $(1,2)$ types. Using the HH algorithm,
we would have to connect both out-stubs of such a node to the nodes with the
largest in-degrees, that is to the two $(3,0)$ types. However, the digraph in
Fig.~\ref{fig:bds1}c) does not have a $(1,2)$ node being connected to both
$(3,0)$ nodes, yet it realizes the sequence. The limitation of the HH algorithm
comes from the fact that it prescribes to connect the out-stub of a node $i$
to an in-stub of the node with the \emph{largest} residual in-degree that does not yet
receive a connection from node $i$. However, there can be other nodes whose in-stubs
can form a connection with an out-stub of $i$ without breaking graphicality.
This shows the importance of finding
a method able to build not just \emph{a} realization of a bds, but \emph{all}
the possible realizations of any given bds.

In the remainder, given a residual bds $\overline{\bm{D}}$,
we denote by ${\cal A}_i\left(\overline{\bm{D}}\right)$ the \emph{allowed set}
of $i$, i.e., the set of all nodes to which an out-stub of
$i$ can be connected without breaking graphicality.
Also, let us denote by ${\cal X}_i\left(\overline{\bm{D}}\right)$ the set of nodes to which
connections were already made from $i$, thus representing the star constraint 
at that stage.

The graphicality test under a star constraint on node $i$ is provided
as Theorem~\ref{star} below. In order to announce it, however, we need
to introduce one more definition.
Consider a bds $\bm{D}$ and a given node $i$ with out-degree $d^{(o)}_i > 0$
from this bds. Let us also consider a subset of nodes $S \subset V$ such that
$|S| \leq d^{(o)}_i$, where $|S|$ denotes the number of nodes in $S$, i.e., its size,
and for every node $j \in S, d^{(i)}_j > 0$. Next, we take $\bm{D}$  and reduce
by unity the in-degrees of all its nodes in $S$, then reduce by $|S|$ the out-degree
of node $i$. The bds $\bm{D'}$ thus obtained will be called the bds
\emph{reduced by $S$ about node $i$ from bds $\bm{D}$}. Equivalently, $\bm{D'}$
is the residual sequence obtained from $\bm{D}$ after connecting an out-stub from
$i$ to an in-stub of every node from $S$.
\begin{theo}[Star-constrained graphicality]\label{star}
Let $\bm{D}$ be a bds in normal order on $N$ nodes,  and let ${\cal X}_i$,
$|{\cal X}_i| \leq N-1 - d^{(o)}_i$, be a set of nodes
whose in-stubs are forbidden to be connected to the out-stubs of node $i$
(including $i$).
Define ${\cal L}_i$ as the set of the first ("leftmost") $d^{(o)}_i$  nodes in $\bm{D}$
but not from ${\cal X}_i$. Then, there exists a simple
digraph which realizes $\bm{D}$ and avoids connections from
$i$ to ${\cal X}_i$, if and only if the bds $\bm{D'}$ reduced by ${\cal L}_i$
about node $i$ from $\bm{D}$ is graphical.
\end{theo}
The proof of this theorem is found in Ref.~\cite{Erd09}. What this theorem does
is to turn a star-constrained graphicality problem for bds $\bm{D}$ into an
\emph{unconstrained one} on the reduced bds $\bm{D'}$. The graphicality of $\bm{D'}$
is then easily tested via the FR theorem. The set ${\cal L}_i$ as defined above
will be called the \emph{leftmost} set for node $i$.

Although announced in its full generality, as ${\cal X}_i$ could be any
predefined subset of nodes with $|{\cal X}_i| \leq N-1 - d^{(o)}_i$, this theorem
applies directly to the digraph construction process when ${\cal X}_i$ represents
the set of nodes to which connections were already made in previous steps
from the same node $i$, hence forbidding us to make further connections from $i$
to these very same nodes. In this case, the bds $\bm{D}$ represents the
residual sequence $\overline{\bm{D}}$ at that stage of the construction process.

As discussed above, in order for us to be able to construct all the simple digraphs
that realize a given bds, we need to find the allowed set
${\cal A}_i\left(\overline{\bm{D}}\right)$ for the next out-stub of $i$.
Clearly, after every connection from the same node $i$,
the residual sequence changes, and along with it the allowed set may change as well.
In order to find ${\cal A}_i\left(\overline{\bm{D}}\right)$ for the next out-stub of node $i$,
we could just simply attempt connections sequentially to every node with non-zero
in-degree \emph{not in} ${\cal X}_i\left(\overline{\bm{D}}\right) 
\cup \left\lbrace i\right\rbrace$, and test for graphicality after each attempt 
using Th.~\ref{star}. The set of nodes for which graphicality would have been
preserved would form ${\cal A}_i\left(\overline{\bm{D}}\right)$.

However, this would be inefficient and, actually, not needed.
In fact, we can exploit a result which states that, if graphicality is broken 
by a connection,
it will be broken by all other connections to the right of the previous one,
in the normal order sense. This is expressed in the following:
\begin{theo} \label{max}
Let $\bm{D}$ be a graphical bds
in normal order and let ${\cal X}_i $ be a forbidden set for node $i$, with 
$i \in {\cal X}_i$. Let $j < k$ be two
nodes such that $j,k\notin{\cal X}_i$. If the residual bds $\bm{D}_j$ obtained
from $\bm{D}$ after forming an edge directed from $i$ to $j$ is not graphical,
then the bi-degree sequence $\bm{D}_k$ obtained from $\bm{D}$ by forming
a directed edge from $i$ to $k$ is also not graphical.
\end{theo}
This theorem follows from the direct contraposition of Lemma 6 in Ref.~\cite{Erd09}.
Thus, what we need to do is to find efficiently the \emph{leftmost node} $q$ in the
residual sequence in normal order, a connection to which would break graphicality.
We will refer to this node $q$ as the {\em leftmost fail-node}. 
All connections to this node and to nodes to its right
are guaranteed to break (star-constrained) graphicality, whereas all connections to
its left (with the exception of forbidden nodes and self) are
guaranteed to preserve the graphical character.

Note that both theorems~\ref{star} and~\ref{max} are based on the HH theorem for
bi-degree sequences. In fact theorem~\ref{star} is a generalization of the HH theorem
to include star constraints. Also note that, while for the FR theorem only the 
in-degrees must be ordered
non-increasingly, for the HH theorem and hence for both theorems \ref{star} and \ref{max},
the bds must be in normal order, as ordering by in-degrees only is not sufficient. This is
easily seen from the following example of graphical bds (not in normal order)
$\bm{D}_7=\left\lbrace(2,0), (2,1), (0,1), (0,2)\right\rbrace$. Using the HH theorem, if we do not
worry about normal ordering, but just order by in-degree, we could choose
to connect the out-stub of node $(0,1)$ to an in-stub of node $(2,0)$, then the out-stub
of node $(2,1)$ to the remaining in-stub of $(2,0)$ (connecting to the largest residual 
allowed residual in-degree), after which we have clearly broken graphicality: both 
out-stubs of $(0,2)$ now must be connected to the two in-stubs of $(2,1)$. 

We are now ready to present our digraph construction algorithm, which produces
random samples from the set of all \emph{possible} simple digraphs realizing a given bds.

\section{The algorithm}\label{thealg}
Given a graphical bi-degree sequence $\overline{\bm{D}}$ in \emph{normal order} (initially
$\overline{\bm{D}} = \bm{D}$):
\begin{my_enumerate}
\item[1)] Define as \emph{work-node} the lowest-index node $i$ with non-zero
(residual) out-degree.
\item[2)] Let ${\cal X}_i$ be the set of forbidden nodes for the work-node,
which includes $i$, nodes with zero in-degrees
and nodes to which connections were made from $i$, previously.
In the beginning, $\mathcal X_i$ includes only the work-node and zero in-degree nodes.
\item[3)] Find the set of nodes, ${\cal A}_i$ that can be connected to the work-node
without breaking graphicality.
\item[4)] Choose a node $m \in {\cal A}_i$ uniformly at random and connect an
out-stub of $i$ to an in-stub of $m$.
\item[5)] After this connection add  node $m$ to ${\cal X}_i$.
\item[6)] If node $i$ still has out-stubs, bring the residual sequence in normal order, then
repeat the procedure from 3) until all out-stubs of the work node are connected away into edges.
\item[7)] If there are other nodes left with out-stubs, reorder the residual degree sequence
in normal order, and repeat from 1).
\end{my_enumerate}
The most involved step of the algorithm is finding the allowed set (step 3)), which is described next.

\subsection{Finding the allowed set}\label{derivations}

Let $i$ be the work-node chosen as in 1) and let $\overline{\bm{D}}$ denote the 
{\em normal ordered},
residual sequence obtained after having connected some of the out-stubs of $i$ to in-stubs
of other nodes, such that graphicality is still preserved. These previous connections
from node $i$ form the set of forbidden nodes ${\cal X}_i$ for the next out-stub
$\sigma$ of $i$. $\mathcal X_i$ also contains the work-node $i$ itself $i \in {\cal X}_i$
and all other nodes with zero in-degrees.
Let ${\cal L}_i$ be the set of the first (lowest index) $\overline{d}^{(o)}_i$ nodes
from $\overline{\bm{D}}$, \emph{not in} ${\cal X}_i$.
As $\overline{\bm{D}}$ is (star constrained) graphical, we can connect
$\sigma$ to any of the nodes in ${\cal L}_i$ without breaking graphicality (due to Theorem \ref{star}),
hence ${\cal L}_i \subseteq {\cal A}_i$.

Let $m$ be the last element of ${\cal L}_i$ in the normal ordered bds
$\overline{\bm{D}}$ and let us "color" (label) red all the non-forbidden nodes,
i.e., all the nodes not in ${\cal X}_i$, to the right of node $m$. Please note that
these color labels are associated with the nodes, defined by their bi-degrees, and not
with their indices of location in the sequence. This set of red nodes ${\cal R}_i$ forms 
the set of candidates for the leftmost fail-node $q$. All other nodes are colored (labelled) black.
To find the leftmost fail-node we could simply connect out-stub $\sigma$ to an in-stub of a red
node $\ell$, add the new connection temporarily to the set of forbidden
nodes, bring the new residual
sequence into normal order, then test for graphicality using Theorem~\ref{star}. 
This procedure could then be repeated sequentially, with $\ell$ going
over all the red nodes from left to right, until graphicality would fail for the first
time at $\ell = q$. However, the considerations in the following paragraphs
allow us define a better method.

For the sake of argument let us perform the sequential testing as explained above. It 
would imply the following steps for a given red node $\ell$ :
\begin{my_enumerate}
\item[(a)] Reduce the out-degree at the work-node $i$ and the in-degree at $\ell$ by unity,
that is $\overline{d}^{(o)}_i \mapsto \overline{d}^{(o)}_i - 1$ and 
$\overline{d}^{(i)}_{\ell} \mapsto \overline{d}^{(i)}_{\ell} - 1$, resulting in a new residual
bds $\overline{\bm{D}}_{\ell}$.
\item[(b)] Bring $\overline{\bm{D}}_{\ell}$ into normal order (required by Theorem 2).
Note that $\ell$ is the only node whose in-degree has changed and only the 
work-node had its out-degree changed (its in-degree was not affected). Thus, when 
bringing $\overline{\bm{D}}_{\ell}$ into normal order, the relative positioning of all the other nodes
does not change. The work-node might have shifted to the right to a new position $i'$
within the block of nodes with the {\em same} in-degree ($i' \geq i$), 
and the red node's new position $\ell'$
might have also moved to the right in the normal ordered sequence ($\ell' \geq \ell$).  
\item[(c)] Add $\ell'$ to the forbidden set for the work-node. 
\item[(d)] Now, as required by Theorem~\ref{star}, reduce by unity the in-degrees of 
the nodes in the left-most adjacency set ${\cal L}_{i'}$, and reduce the 
out-degree of the work-node $i'$ to zero. 
This results in the new sequence  $\overline{\bm{D}'}_{\ell'}$.
\item[(e)] Order the bds $\overline{\bm{D}'}_{\ell'}$ by in-degrees, non-increasingly.
\item[(f)] Apply the FR theorem to test for graphicality. 
\end{my_enumerate}
Thus, whether the connection of the work-node $i$ to $\ell$ breaks graphicality, 
ultimately depends on whether the residual bds $\overline{\bm{D}'}_{\ell'}$ fails (or passes) the
FR test. However, as we noted before, for the FR test 
we do not need to have the bds $\overline{\bm{D}'}_{\ell'}$ in normal order, we
only need to have it ordered non-increasingly by the in-degrees. Additionally, observe
that in step (d)  the reduction of the in-degrees always happens on the {\em same} set of
nodes, independently of the red node $\ell$, that is the left-most adjacency set ${\cal L}_{i'}$
is the same for all $\ell$. Thus, in this particular case of  Theorem 2's application, ultimately we
do not need to bring  $\overline{\bm{D}}_{\ell}$ into normal order (step (b)), only non-increasingly
by in-degrees, which would be done anyway in step (e). 
That means we can just skip step (b), we do not need to move around any of the nodes at that stage.
Thus, the only difference between the sequences $\overline{\bm{D}'}_{\ell'}$ for different
$\ell$-s is at the position of this node after the reordering in (e), {\em with respect to the rest of the 
sequence}. 

These observations suggest that we should define a bds $\overline{\bm{D}'}$ obtained
from the bds $\overline{\bm{D}}$ by reducing by unity the in-degrees of all nodes in the 
set ${\cal L}_{i}\setminus \{m\}$ and by $\overline{d}^{(o)}_i - 1$  the out-degree of the 
work-node $i$, leaving only one out-stub (out-stub $\sigma$) at $i$. 
Clearly, the bds $\overline{\bm{D}'}$ is graphical (connecting out-stub $\sigma$ to an 
in-stub of node $m$ surely preserves graphicality, by Theorem \ref{star}). 
Let us now order $\overline{\bm{D}'}$ non-increasingly by its in-degrees, {\em in a specific
way}, described as follows. Shift only the reduced in-degree nodes in $\overline{\bm{D}}$ 
to the right with respect to the rest of the sequence such as to restore non-increasing ordering 
by the in-degrees (if needed).
Since only the in-degrees of the nodes in the set ${\cal L}_{i}\setminus \{m\}$ have been
reduced, {\em keep} the relative ordering of all other nodes in $\overline{\bm{D}'}$ 
exactly the same
as in $\overline{\bm{D}}$. Thus the {\em relative ordering} of the red nodes and of the 
work node have been preserved as well. Let us denote the new location of the work node in $\overline{\bm{D}'}$ by $j$
($j \leq i$).
Connecting now $\sigma$ to an in-stub of a red node $\ell$ in this sequence 
will produce the same set of residual bi-degrees as in step (d) above.
To be able apply the FR theorem, then all we need to do is to shift to the right node $\ell$ in 
the sequence (if needed) to make sure that it is non-increasingly ordered by in-degrees. 
Since only the in-degree at $\ell$ was modified (reduced by unity), this reordering is very 
simple: if $x$ denotes the location of the 
last node of the block of nodes with the same in-degree as 
node $\ell$ in $\overline{\bm{D}'}$  ($x \geq \ell$), then we simply swap the node at 
$\ell$ with the node at $x$ after the reduction of the in-degree at $\ell$.   Let us denote the obtained  
sequence  by $\overline{\bm{D}''}$. Clearly, it is non-increasingly ordered by in-degrees, and
thus we can apply the FR theorem to see if it is graphical. Note: it could happen that 
$x=j$ (e.g., there are many nodes with zero out-degree but larger in-degree than the work-node
as defined in 1)), however, the steps below can be applied just the same.

Next, we show how to identify the leftmost red fail-node $q$ by
investigating how the inequalities in~(\ref{FReq}) break down.
Since $\overline{\bm{D}'}$ is graphical, we have for all $1\leq k \leq n-1$
($n$ is the last element of $\overline{\bm{D}'}$) that $L'(k) \leq R'(k)$,
where $L'$ and $R'$ are the left hand side (lhs) and the right hand side (rhs)
of inequalities~(\ref{FReq}) written for $\overline{\bm{D}'}$:
\begin{eqnarray}
L'(k) &=& \sum_{s=1}^{k} \overline{d'}^{(i)}_s\;, \label{insL} \\
R'(k) &=& \sum_{s=1}^{k}\min\left\lbrace k-1,
\overline{d'}^{(o)}_s\right\rbrace
\!+\!\!\!\! \sum_{s=k+1}^{n}\!\!\!\min\left\lbrace k, \overline{d'}^{(o)}_s\right\rbrace. \label{insR}
\end{eqnarray}
Let us denote by $L''(k)$ and $R''(k)$ the lhs and rhs
of the inequality~(\ref{FReq}) corresponding to $\overline{\bm{D}''}$.
Since the rhs of~(\ref{FReq}) involves only out-degrees, and we only reduced the
out-degree of the work-node from 1 to 0, we will always have $R''(k) = R'(k)-1$,
\emph{except} when  $k=1$ and the work-node is at $j=1$, in which case $R''(1) = R'(1)$. 
However,
in this case, $L''(1) = L'(1)$, because only the in-degree of $j=1$ appears, which
does not get changed. Thus, since $L'(1) \leq  R'(1)$ ($\overline{\bm{D}'}$ 
is graphical), graphicality cannot be broken at $k=1$ when $j=1$. Let us now 
consider that the work-node is still at position $j=1$, but $k > 1$. 
For $1 < k < x$, the in-degrees in
$\overline{\bm{D}''}$ are the same as those in $\overline{\bm{D}'}$, hence
$L''(k) = L'(k)$. For $k  \geq x$, however, we have $L''(k) = L'(k)-1$.
Now consider $j > 1$.  For  $1 \leq k < x$, we have 
$L''(k) = L'(k)$ and for $k \geq x$, $L''(k) = L'(k)-1$. The following summarizes 
the relationships above:
\begin{my_enumerate}
\item[(A)] $j=1$:
\begin{my_enumerate}
\item[(A.1)] $k=1$:
$\qquad\;\;L''(1) = L'(1)\;,\quad\;\;\;\;\;\;\;\;R''(1) = R'(1)\;.$
\item[(A.2)] $1 < k < x$: 
$\;\;\;\;L''(k) = L'(k)\;,\quad\;\;\;\;\;\;\;\;R''(k) = R'(k)-1\;.$
\item[(A.3)] $x \leq k$:
$\quad\;\;\;\;\;\;\; L''(k) = L'(k)-1\;,\quad\;\;R''(k) = R'(k)-1\;.$
\end{my_enumerate}
\item[(B)] $j>1$:
\begin{my_enumerate}
\item[(B.1)] $1 \leq k < x$:
$\;\;\;\;L''(k) = L'(k)\;,\quad\;\;\;\;\;\;\;\;R''(k) = R'(k)-1\;.$
\item[(B.2)] $x \leq k$:
$\quad\;\;\;\;\;\;\; L''(k) = L'(k)-1\;,\quad\;\;R''(k) = R'(k)-1\;.$
\end{my_enumerate}
\end{my_enumerate}
Since $L'(k) \leq  R'(k)$ for all $k$, 
graphicality for $\overline{\bm{D}''}$ can only be broken (that is to have $L'' > R''$ for some $k$),
if  $L'(k) = R'(k)$, namely in cases (A.2) and (B.1) above.
Observe that $L'(k)$ and $R'(k)$ are computed from $\overline{\bm{D}'}$, hence
they are independent from $\ell$ or $x$. This gives us the following
simple procedure for finding the leftmost fail-node, if it exists.
Starting from $k=2$ for $j=1$, and $k=1$ for $j > 1$, find the
smallest $k_0$ for which $L'(k_0) = R'(k_0)$. If no such $k_0$ exists, then there
are no fail-nodes and all non-forbidden nodes are to be included in the allowed set.
If there is such a $k_0$, the first red node $q'$ to the right of $k_0$ ($q' \geq
k_0+1$) is the leftmost fail-node of $\overline{\bm{D}''}$, which when identified
in the original bds $\overline{\bm{D}}$ will give the leftmost fail-node $q$.
All non-forbidden nodes to the left of $q$ are to be included in the allowed set.

\subsection{Summary for finding the allowed set}\label{summary}

What we discussed in detail in the previous subsection corresponds to step (3)
of the main algorithm described in the beginning of Section~\ref{thealg}.
Given the normal-ordered bds $\overline{\bm{D}}$ at the end of step 2)
of the main algorithm:
\begin{my_enumerate}
\item[(3.1)] Identify ${\cal L}_i$ from the first $\overline{d}^{(o)}_i$ nodes not in ${\cal X}_i$.
\item[(3.2)] Identify the ``red'' set ${\cal R}_i$ as those nodes that
are neither in ${\cal L}_i$ nor in ${\cal X}_i$. Note, the color label is associated with 
the node, not its index.
\item[(3.3)] Build $\overline{\bm{D}'} $ as follows:
\begin{displaymath}
 \overline{d'}^{(i)}_b = \left\lbrace
 \begin{array}{ll}
 \overline{d}^{(i)}_b -1  & \textrm{if $b \in {\cal L}_i \setminus \left\lbrace m\right\rbrace$}\\
 \overline{d}^{(i)}_b  & \textrm{otherwise}
  \end{array} 
\right.
\end{displaymath}
and
\begin{displaymath}
 \overline{d'}^{(o)}_c = \left\lbrace
 \begin{array}{ll}
 1  & \textrm{if $c = i$}\\
 \overline{d}^{(o)}_c  & \textrm{otherwise}
  \end{array} 
\right.
\end{displaymath}
where $m$ is the last node in $\mathcal L_i$.
\item[(3.4)] Shift nodes from ${\cal L}_i \setminus \left\lbrace m\right\rbrace$ to the right 
in the sequence (and only these) such as to restore ordering non-increasingly by in-degrees
(if needed), preserving the color labels of the nodes in the process.
The work-node may have shifted to a new location $j$ after this step. This is the updated
sequence $\overline{\bm{D}'} $.
\item[(3.5)] 
Starting from $k =1$ if $j \ne 1$ or from $k = 2$ if $j = 1$, find
$k_0$ as the smallest $k$ such that $L'(k) = R'(k)$, where $L'(k)$ and $R'(k)$
are computed from {\em the reordered} (after step (3.4)) $\overline{\bm{D}'}$ using 
(\ref{insL}) and (\ref{insR}). If there is no
such $k_0$, then the allowed set ${\cal A}_i$ is all the nodes in
$\overline{\bm{D}}$ except nodes from the forbidden set ${\cal X}_i$.
\item[(3.6)] Otherwise find the leftmost red node $q'$ in the updated
bds $\overline{\bm{D}'}$
to the right of $k_0$, that is with $q' > k_0$. Then the corresponding node $q$
in $\overline{\bm{D}}$ , will be the leftmost fail node. Note that 
$q'$ is the new position of the node at $q$ in $\overline{\bm{D}}$ after
the reordering in (3.4).
\item[(3.7)] The allowed set ${\cal A}_i$ is formed by all nodes
in $\overline{\bm{D}}$ not in ${\cal X}_i$, and to the left of $q$.
\end{my_enumerate}

\section{The sampling problem}\label{samprob}

The algorithm generates an independent sample digraph every time it runs,
\emph{without restarts or rejections,} and it guarantees that \emph{every possible}
realization of a graphical bds by simple digraphs can be generated with a
non-zero probability. However, the algorithm realizes the digraphs
with non-uniform probability. Nevertheless, knowing the relative probability for
every digraph's occurrence allows us to calculate network observable
averages as if they were obtained from a uniform sampling.
In particular, the following expression, which is a well-known result in biased
sampling~\cite{Coc77,New99}, provides these averages as:
\begin{equation}\label{two}
\langle Q\rangle=\frac{\sum_{j=1}^M w(\bm{s}_j) Q(\bm{s}_j)}
{\sum_{j=1}^{M}w(\bm{s}_j)}\:,
\end{equation}
where $Q$ is an observable measured from the samples $\bm{s}_j$ generated by an algorithm.
The $w(\bm{s}_j)$ sample weight is the inverse of the relative probability of the occurrence
of $\bm{s}_j$ and $M$ is the number of the samples generated. In
Subsection~\ref{sec:bclass} we give a detailed derivation of this formula, specialized to our graph
construction problem. The weights of the samples generated by our algorithm are given by
\begin{equation}\label{weight}
w(\bm{s})=\prod_{i}\prod_{j=1}^{d^{(o)}_i}k_{i}(j)\;.
\end{equation}
where $i$ runs over all the nodes with non-zero out-degree
as they are picked by the algorithm
to become work-nodes, and $k_{i}(j) = |{\cal A}_i(j)|$ is the size of the allowed
sets ${\cal A}_i(j)$ just before connecting the $j$-th out-stub of $i$.
Note that $w \geq 1$ since there always exists at least one digraph realizing the bds.
Subsection~\ref{sec:weights} gives a derivation of~(\ref{weight}).

\subsection{Biased sampling over classes}\label{sec:bclass}

Our algorithm sequentially connects all stubs from a series of work nodes and finishes with
a simple, labeled digraph. This process can be uniquely described by a \emph{path} of
connection sequences. Having chosen a work node $i_1$ for the first time, it
determines the allowed
set ${\cal A}_{i_1}$. We next choose uniformly at random a node
$j_1(i_1) \in {\cal A}_{i_1}$ and connect a stub of $i_1$ to a stub at
$j_1(i_1)$. We could have chosen $j_1(i_1)$
following any other criterion, but in that
case the expression~(\ref{weight}) of the weights would have to be modified accordingly.
After this connection we recompute the new allowed set
${\cal A}_{j_1}(i_1)$, then connect another stub of $i_1$, and so on until all the stubs
have been used up at $i_1$. Let us denote by $\bm{s}$ such a path of connection
sequences:
\begin{equation}
\bm{s} = \left\lbrace i_1,j_1(i_1),\ldots,j_{\bar{d}^{(o)}_{i_1}}(i_1);
i_2,j_2(i_2),\ldots, j_{\bar{d}^{(o)}_{i_2}}(i_2)\ldots \right\rbrace \label{path}
\end{equation}
where $\bar{d}^{(o)}_i$ denotes the residual out-degree of node $i$. A path $\bm{s}$ uniquely defines
the digraph $\bm{G}(\bm{s})$ created, as the collection of all connections in~(\ref{path}) forms the
edge set of the created graph $\bm{G}(\bm{s})$. However, several paths may lead to the same digraph.
Also note that the order of the connections in~(\ref{path}) matters in the calculation of the weight, as the
corresponding allowed sets in general depend on history of connections up to that point.
For a finite bi-degree sequence
the number of distinct samples (paths) is also finite. Let us denote this set of paths by:
\begin{equation*}
\Pi = \left\lbrace \bm{\pi}_1,\ldots,\bm{\pi}_P\right\rbrace\;.
\end{equation*}
Let us now assume that we built with our algorithm a sequence of
samples $\bm{s}_1,\bm{s}_2,\ldots,\bm{s}_M$, and that the sample number $M$
is large enough for us to see all elements of $\Pi$ sufficiently many times. Given some path
$\bm{s}$ we compute a quantity $Q(\bm{s})$, and we are interested in
calculating the average of $Q$ over path ensembles. In our case $Q$ is defined on the
final graph itself $Q(\bm{s}) = Q(\bm{G}(\bm{s}))$, but for now we will not consider that,  explicitly. If we
were just simply computing the average of $Q$ over the set of samples, we would
obtain an average \emph{biased} by the way the algorithm
builds the paths from $\Pi$:
\begin{equation}
\langle Q \rangle = \frac{1}{M}\sum_{i=1}^{M} Q(\bm{s}_i) = \sum_{k=1}^{P} 
\frac{M_k}{M} Q(\bm{\pi}_k)\;, \label{direct}
\end{equation}
where $M_k$ is the number of times we have seen path $\bm{\pi}_k$ appear in the sequence
of samples. Clearly,
\begin{equation}
\rho_k = \lim_{M\to \infty} \frac{M_k}{M} \label{rhos}
\end{equation}
is the probability by which path $\bm{\pi}_k$ is generated via the algorithm. We
now assume that we can compute analytically the path probabilities $\rho_k$, from knowing
how the algorithm works. Instead of~(\ref{direct}) we want to compute the average as if it
was measured over the
uniform ensemble of paths, that is:
\begin{equation}
\langle Q \rangle_{up} = \frac{1}{P}
\sum_{k=1}^{P} Q(\bm{\pi}_k)\;. \label{unif}
\end{equation}
If we form:
\begin{eqnarray}
\langle Q \rangle_{bp} &=& \frac{\sum_{i=1}^{M} \frac{1}{\rho(\bm{s}_i)} Q(\bm{s}_i)}
{\sum_{i=1}^{M} \frac{1}{\rho(\bm{s}_i)}}  \label{biasedw}\\
&=& \frac{\sum_{k=1}^{P} \frac{M_k}{M\rho(\bm{\pi}_k)} Q(\bm{\pi}_k)}
{\sum_{k=1}^{P} \frac{M_k}{M\rho(\bm{\pi}_k)}}\;,  \nonumber
\end{eqnarray}
we have $\lim_{M \to \infty} \langle Q \rangle_{bp}   = \langle Q \rangle_{up}$,
due to~(\ref{rhos}). Thus, the weighted average~(\ref{biasedw}) should be used
in order to obtain averages according to uniform sampling in the $M \gg 1$ limit.

Let us assume that there is an equivalence relation "$\sim$" between paths,
hence inducing a partitioning of  $\Pi$ into $K$ equivalence classes: $\Pi = {\cal C}_1
\cup \ldots \cup {\cal C}_K$, where
${\cal C}_{\ell} = \left\lbrace\bm{\pi}_{k_1^{\ell}},\ldots,\bm{\pi}_{k_{\mu_{\ell}}^{\ell}}\right\rbrace$.
The size of class ${\cal C}_{\ell}$ is denoted by $\mu_{\ell} = \left| {\cal C}_{\ell}\right|$.
We have $\sum_{\ell = 1}^K \mu_{\ell} =P$.
Alternatively, for some given path $\bm{\pi}$, we will denote by ${\cal C}(\bm{\pi})$
the equivalence class of  $\bm{\pi}$ and by $\mu(\bm{\pi}) = 
\left| {\cal C}(\bm{\pi}) \right|$ its size.
Let us also assume that if $\bm{s},\bm{r} \in {\cal C}_{\ell}$,
that is $\bm{s} \sim \bm{r}$, then $Q(\bm{s}) = Q(\bm{r})$.
For example, in our case distinct paths may lead to the
same digraph. We introduce
the equivalence relation "$\sim$" and say that two paths $\bm{s}$ and $\bm{r}$
are equivalent, $\bm{s} \sim \bm{r}$ if they produce the same labeled digraph,
$\bm{G}(\bm{s}) = \bm{G}(\bm{r})$. Clearly, if $Q$ depends only on the
constructed graph, i.e.,
$Q(\bm{\pi})=Q\left(\bm{G}(\bm{\pi})\right)$ for all $\pi \in \Pi$, then $Q(\bm{s}) = Q(\bm{r})$
whenever $\bm{s} \sim \bm{r}$.

Our goal is to obtain the average of $Q$ uniformly over the equivalence
classes, that is:
\begin{equation}
\langle Q \rangle_{uc} = \frac{1}{K}
\sum_{\ell=1}^{K} Q\left(\bm{\pi}_{k_1^{\ell}}\right)\;, \label{unicl}
\end{equation}
where we chose to write the first element of ${\cal C}_{\ell}$ in the argument of
$Q$, but of course, any other element could have been chosen from the same
class, as $Q$ is constant within a class. In general, (\ref{biasedw}) will not
produce $\langle Q \rangle_{uc}$, but a sum weighted by class sizes. Instead, let us
consider:
\begin{equation}
\langle Q \rangle_{bc} = \frac{\sum_{i=1}^{M} \frac{1}{\mu(\bm{s}_i)\rho(\bm{s}_i)} Q(\bm{s}_i)}
{\sum_{i=1}^{M} \frac{1}{\mu(\bm{s}_i)\rho(\bm{s}_i)}} \;. \label{biasedcw}
\end{equation}
It is then easy to see that:
\begin{equation*}
\langle Q \rangle_{bc}  =  \frac{\sum_{k=1}^{P} \frac{M_k/M}
{\mu(\bm{\pi}_k)\rho(\bm{\pi}_k)} Q(\bm{\pi}_k)}
{\sum_{k=1}^{P} \frac{M_k/M}{\mu(\bm{\pi}_k)\rho(\bm{\pi}_k)}} 
\stackrel{M\to\infty}{\longrightarrow} \langle Q \rangle_{uc}\;.
\end{equation*}
In order for (\ref{biasedcw}) to be useful in practice, one has to be able to compute
the size of the equivalence class $\mu(\bm{s})$ from seeing $\bm{s}$
and knowing how the algorithm works.
Fortunately this is possible in our case, as shown next.

\subsection{Computing the weights}\label{sec:weights}

First, let us note that when connecting the out-stubs of a work-node
we are not affecting the out-stubs of any other nodes, but only in-stubs. 
Hence, all nodes with non-zero out-degrees will eventually be picked
as work-nodes by the algorithm. Since normal ordering is first by
in-degrees, the \emph{order} in which nodes will become work-nodes
depends on the sequence of connections. Let us now calculate the
probability of the path $\bm{s}$ in (\ref{path}). Given a residual sequence,
the work-node $i_1$ is uniquely determined by the algorithm as described
before. Since the next connection is picked uniformly at random,
the probability of the link from $i_1$ to
$j_1(i_1) \in {\cal A}_{i_1}(j_1)$ is $|{\cal A}_{i_1}(j_1)|^{-1}$.
Let $k_{i}(j) = |{\cal A}_{i}(j)|$ denote the number of nodes in
${\cal A}_{i}(j)$. Then, it is easy to see that the probability of a
path $\bm{s}$ is given by:
\begin{equation}
\rho(\bm{s}) = \left[ \prod_{k}\prod_{j=1}^{d^{(o)}_{i_k}}
k_{i_k}(j) \right]^{-1} \label{pathprob}
\end{equation}
where $i_1,i_2,\ldots,$ denote the work-nodes in the order in which they
are picked by the algorithm. This expression can be computed readily
in a computer as the algorithm progresses. In order for us to use
(\ref{biasedcw}) it seems that we would need also to obtain the size
$\mu(\bm{s})$ of the class to which path $\bm{s}$ belongs. Clearly,
two different paths $\bm{s}$ and $\bm{s'}$ will result in the same graph
($\bm{s} \sim \bm{s'}$)
if and only if the sequence of connections in one path is a permutation
of the connections in the other path. Hence, the class size $\mu(\bm{s})$
is nothing but the number of permutations of the connections, which is
the same for all paths, that is, all classes have the same size
$\mu$. Since all connections are made from a node first before moving
on to another, we have $\mu = \prod_{i=1}^{N} d_{i}^{(o)}!$ . However, we
actually don't need to use this number: one can simply multiply by
$\mu$ both the numerator and the denominator of~(\ref{biasedcw}) to
obtain~(\ref{two}-\ref{weight}).

\subsection{A simple example}\label{sec:ex}

In this subsection we illustrate the algorithm on a simple sequence:
$\bm{D}_8=\left\lbrace(2,2), (2,1), (1,3), (1,1), (1,0)\right\rbrace$. There are
11 distinct labeled digraphs realizing this sequence and there
are $2!1!3!1!0! = 12$ paths in a class, leading to the same graph.
Two paths that lead to different graphs are for example
$\bm{s}_1 = \left\lbrace(1,4) (1,2); (3,1), (3,5), (3,2); (2,1); (4,3)\right\rbrace$
(connect an out-stub of node 1 to an in-stub of node 4, etc.)
and $\bm{s}_2 = \left\lbrace(1,2) (1,3); (2,1); (4,1); (3,4), (3,5), (3,2)\right\rbrace$.
For the former, $w(\bm{s}_1)=[\rho(\bm{s}_1)]^{-1} = 8 $ and
for the latter it is $w(\bm{s}_2)=[\rho(\bm{s}_2)]^{-1} = 54$. Let us now
consider the Pearson coefficient $r$ of degree-degree correlations,
or the assortativity coefficient defined for directed graphs~\cite{New03}
as our network
observable $Q=r$.
\begin{figure}
\centering
\includegraphics[width=1\textwidth]{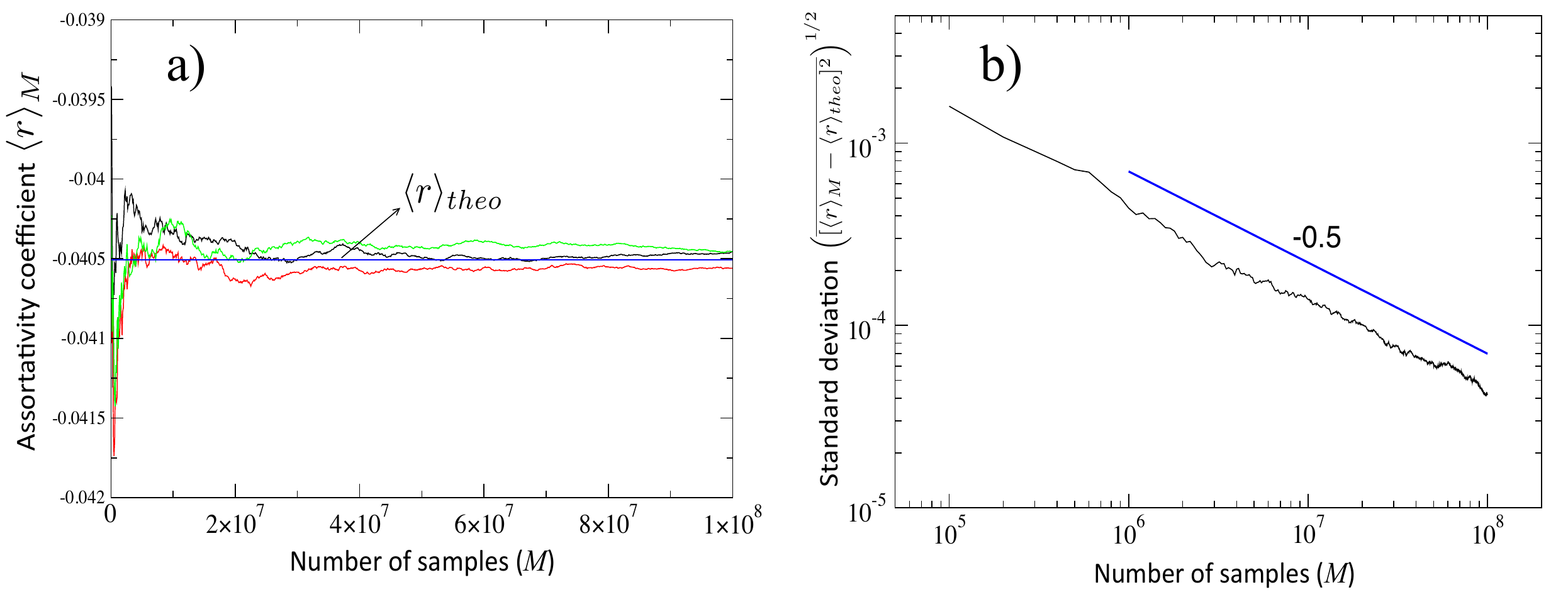}
\caption{Biased sampling on the example bds $\bm{D}_8$.
The measure monitored is Newman's assortativity coefficient $r$~\cite{New03}.
In b) the ensemble average was taken over 50 runs.}\label{fig:simpex}
\end{figure}
For each one of the 11 graphical realizations of $\bm{D}_8, r$
can be calculated exactly, as can the uniform
average over this ensemble, obtaining
$\langle r \rangle_{theo}  = -0.040506$. We will refer to $\langle r \rangle_{theo}$
as the ``theoretical value". We then let our algorithm run on this
sequence to produce $M$ samples
and using~(\ref{two}-\ref{weight}) to obtain the corresponding coefficient
$\langle r \rangle_M$. Fig~\ref{fig:simpex}a) shows a few runs
with different seeds and their convergence to the theoretical value.
Fig~\ref{fig:simpex}b) shows the standard deviation
$( \overline{ [\langle r \rangle_M - \langle r \rangle_{theo}]^2} )^{1/2}$
where the overline denotes an ensemble average over runs.

\section{Complexity of the algorithm}\label{algcomp}

To determine the theoretical upper bound for the
complexity of the algorithm, that is the worst-case complexity, 
notice that there are only three 
steps in the algorithm that require more than $O\left(1\right)$ computational
operations, or steps, to complete.

First, after each connection is placed, one must
bring the residual sequence into normal order, steps 6) or 7).
To accomplish this, both the work-node $i$ and the target node $m$
will have to move to the right, but the relative positions
of all other nodes will remain unchanged. In other words,
if we were to remove nodes $i$ and $m$, the rest of the bds
would already be sorted. Thus, in order to complete these steps, one
only has to find the new positions of nodes $i$ and $m$ and
insert them into the already sorted bds. Therefore, the complexity
of either one of step 6) and step 7) is simply $O\left(2\log N+N\right)
\approx O\left(N\right)$,
where $N$ is the number of the nodes in the sequence being ordered.

Second,
the allowed set $\mathcal A$ must be built before placing each connection (step 3).
Following the summary of this step, given in Subsection~\ref{summary},
notice that steps~3.1 to~3.4 can be all finished during
a single scan of the residual bds. This is clearly so for the creation
of the leftmost set ${\cal L}_i$ and for setting the ``red'' color labels (or flags)
(steps~(3.1) and~(3.4)). Concerning the ordering of the bds $\overline\mathbf{D'}$,
it is possible to create it already sorted by simply scanning the bds
$\overline\mathbf D$ while keeping track of the in-degree $d^\star$ of the
nodes currently being copied and the index $a$ in $\overline\mathbf{D'}$
of the first node with that in-degree. Then, because $\overline\mathbf{D}$
is in normal order, the only possibility for a node in $\overline\mathbf{D'}$
to break the order is if its in-degree equals $d^\star+1$. In this case,
it can be simply swapped with the node at $a$, because, as argued
in Subsection~\ref{derivations}, the mechanism to build the allowed
set is entirely based on the FR theorem, which does not require the
bds to be in normal order, but to be simply ordered non-increasingly
by its in-degrees. Thus, steps~(3.1) to~(3.4) can be completed 
in ${\cal O}\left(N\right)$ steps.

Third, the computation of the sums $L'$, $R'$ and their comparison 
must be conducted, which is the same step as 
(\ref{FReq}) in an FR test.
To determine the complexity of an FR test note that
computing the repeated sums for each one of the inequalities~(\ref{FReq})
is quite inefficient. Instead, below we derive
recurrence relations that allow us to complete the FR test 
in a linear, ${\cal O}\left(N\right)$ number of steps.

The steps of the main algorithm are done sequentially, 
and thus can all be completed in a total of ${\cal O}\left(N\right)$ steps.
They must, however, be repeated for each edge in the digraph.
Thus, the maximum complexity of the algorithm is ${\cal O}\left(N M \right)$
where $M=\sum_i d^{(o)}_i$ is the number of edges. 
Since ${\cal O}\left( M \right) \leq {\cal O}\left(N^2 \right)$
the maximum complexity of the algorithm is $O(N^3)$. It is
important to note though, that for a given
bds the complexity of the algorithm can be substantially smaller, similar to
the case for our undirected graph sampling algorithm~\cite{Del10}.

\subsection{The Fulkerson-Ryser test revisited}

The most complex part of the Fulkerson-Ryser test is to compute
the lhs and the rhs of inequalities~(\ref{FReq}), which we
rewrite here for the sake of readability:
\begin{eqnarray*}
L(k) &=& \sum_{s=1}^{k} d^{(i)}_s\;,\\
R(k) &=& \sum_{s=1}^{k}\min\left\lbrace k-1,
d^{(o)}_s\right\rbrace
\!+\!\!\!\! \sum_{s=k+1}^{N}\!\!\!\min\left\lbrace k, d^{(o)}_s\right\rbrace .
\end{eqnarray*}

Our goal is to find recursion relations for $L(k)$ and $R(k)$. For the lhs
the relation is simply
\begin{equation*}
 L(k+1) = L(k)+d_k^{(i)}\:,
\end{equation*}
with $L(1)=d_1^{(i)}$.

For the rhs, first note that one can write it as
\begin{equation}\label{rhs1}
 R(k) = -k + \sum_{i=1}^N \min\left\lbrace k, g_i^{(o)}(k)\right\rbrace\:,
\end{equation}
where $g_i^{(o)}(k)$ is the family of integer sequences defined as
\begin{equation*}
 g_i^{(o)}(k) = \left\lbrace\stack{d_i^{(o)}+1\quad\forall i\leqslant k}{d_i^{(o)}\quad\forall i>k}\right.\:.
\end{equation*}
Now, let us introduce $G_k\left(p\right) = \sum_{i=1}^{N} \delta_{p,g_i^{(o)}(k)}$, that is, the number of indices $i$ for which $g_i^{(o)}(k)=p$.
Then, from~\eref{rhs1} follows that
\begin{equation}\label{rhs2}
 R(k) = -k + \sum_{p=1}^k pG_k\left(p\right) + k\sum_{p=k+1}^{N}G_k\left(p\right)\:,
\end{equation}
hence
\begin{equation}\label{rhs0}
 R(1) = N-1-G_1\left(0\right)\:,
\end{equation}
where we used the fact that $\sum_{p=0}^{N}G_k\left(p\right)=N$.

Furthermore, let us introduce the following notations:
\begin{eqnarray*}
 \Delta G_k\left(p\right) \equiv G_k\left(p\right) - G_{k-1}\left(p\right)\\
 \tilde G_k\left(q\right) \equiv \sum_{i=0}^q G_k\left(i\right)\:.
\end{eqnarray*}
Then, after some simple manipulations, from~\eref{rhs2} it follows that
\begin{eqnarray}
 \fl R(k) - R(k-1) = N-1-\tilde G_{k-1}\left(k-1\right)\nonumber\\
+\sum_{p=1}^{k-1} p\Delta G_k\left(p\right)+k\sum_{p=k}^{N} \Delta G_k\left(p\right)\:.\label{rhs3}
\end{eqnarray}
Finally, notice that $\Delta G_k\left(p\right) = \delta_{p,d_k^{(o)}+1} - \delta_{p,d_k^{(o)}}$.
Substituting it into~\eref{rhs3}, we obtain:
\begin{eqnarray}
R(k) = \left\{ 
\begin{array}{ll}
R(k-1)+N-\tilde{G}_{k-1} (k-1) & \;\;\forall d_k^{(o)} < k \\
\\
R(k-1)+N-\tilde{G}_{k-1} (k-1) - 1 & \;\;\forall d_k^{(o)} \geqslant k
\end{array} \right. \label{rhsfinal}
\end{eqnarray}

\begin{figure}
\centering
\includegraphics[width=0.7\textwidth]{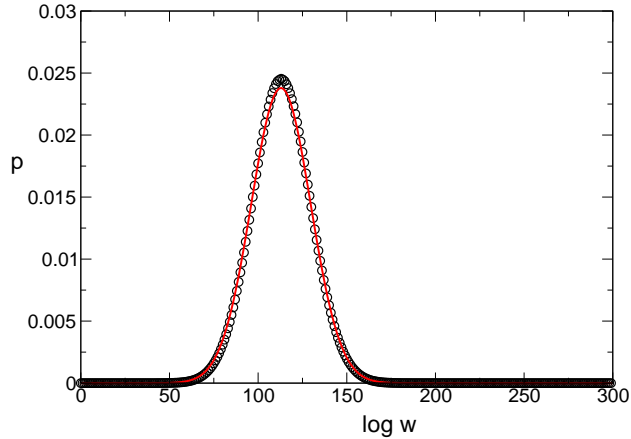}
\caption{Probability distribution $p$ of the logarithm of weights for an ensemble of
bi-degree sequences on $N=100$ nodes. The in-degrees were drawn from a 
normalized power-law distribution $\sim d_{in}^{-\gamma}$ with $\gamma = 3$ 
and the out-degrees were drawn from a Poisson distribution $e^{-\lambda} 
\lambda^{d_{out}}/d_{out}!$, with the same average as the average {\em in-degree},
$\lambda = \langle d_{in}\rangle$. 
The black circles are the simulation data and the red continuous line is a Gaussian
fit. }\label{fig:logweight}
\end{figure}
Thus, we have turned the problem of finding a recursion relation for $R(k)$
into the problem of finding $\tilde G_k\left(k\right)$. To solve this, first
note that
\begin{equation*}
 \tilde G_k\left(k\right) = \tilde G_{k-1}\left(k-1\right)+G_{k-1}\left(k\right)-\delta_{k,d_k^{(o)}}\:,
\end{equation*}
with $\tilde G_1\left(1\right) = G_1\left(0\right) + G_1\left(1\right)$.
The above equation constitutes a recursion relation for $\tilde G_k\left(q\right)$.
Such a relation can be rewritten as
\begin{equation*}
 \tilde G_k\left(k\right) = \tilde G_{k-1}\left(k-1\right)+G_1\left(k\right)+S\left(k\right)\:,
\end{equation*}
where
\begin{equation*}
 S\left(k\right) = \sum_{t=2}^{k-1}\delta_{k,d_t^{(o)}+1} - \sum_{t=2}^k\delta_{k,d_t^{(o)}}\:.
\end{equation*}
Observe that $S\left(k\right)$ and $G_1\left(k\right)$ can be easily computed while scanning the bds,
and then calculating $L(k)$ and $R(k)$ for each $k$ requires a single operation. Thus,
the entire FR test can be completed in $O\left(N\right)$ steps.

\section{Discussion}\label{discussion}
In summary, we have developed a graph construction and sampling algorithm
to construct simple directed graphs realizing a given sequence of in- and out-degrees.
Such constructions are needed in practical modeling situations,
ranging from epidemics and sociology through food-webs to transcriptional
regulatory networks, where we are interested in learning about the statistical
properties of the network observables as determined \emph{only by the
bi-degree sequence} and nothing else.

\begin{figure}
\centering
\includegraphics[width=0.7\textwidth]{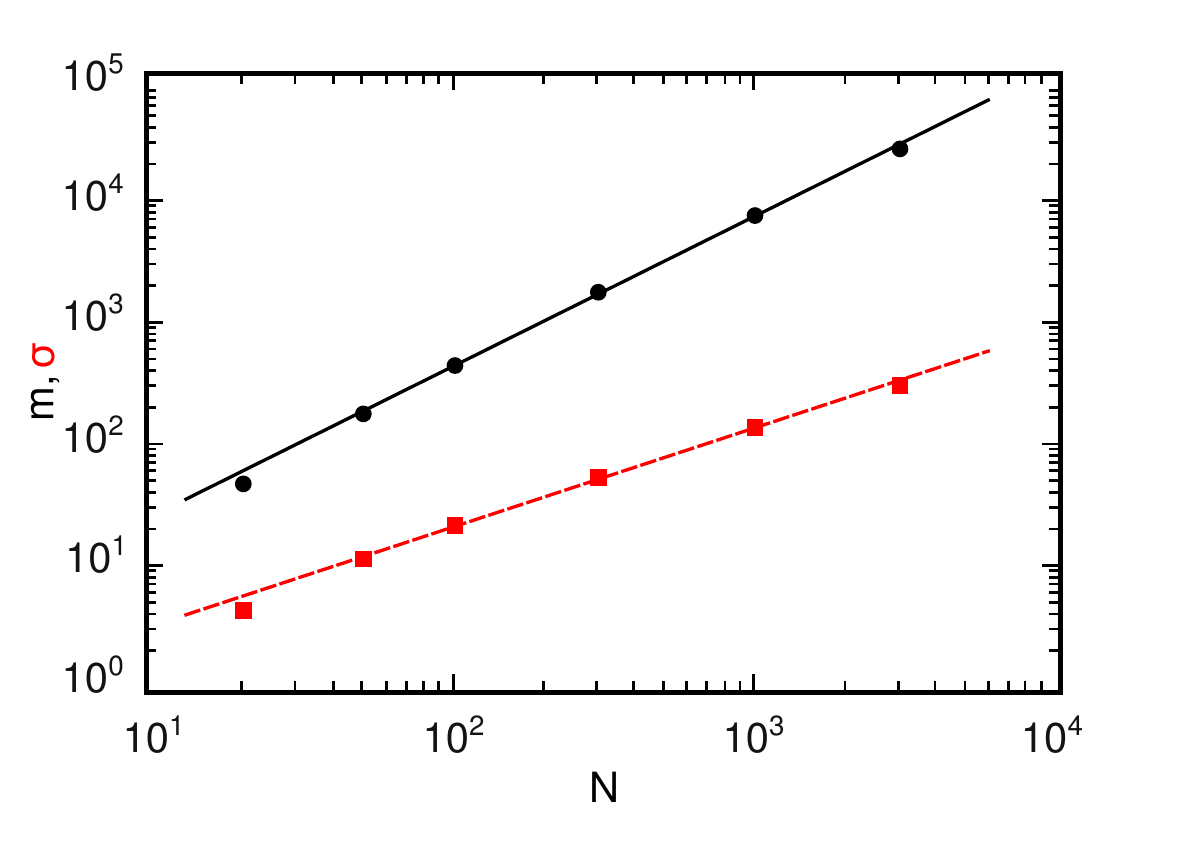}
\caption{Mean $m$ (black circles) and standard deviation $\sigma$ (red squares) of the distributions
of the logarithm of the weights vs.\ number of nodes $N$ of samples. In-degrees and out-degrees are
both drawn from a power-law distribution $P\left(d\right)\sim d^{-\gamma}$, with $\gamma=3$. The solid
black line and the dashed red line are data fit results, showing that $m$ and $\sigma$ follow power-law
scaling laws $m\sim N^\alpha$ and $\sigma\sim N^\beta$. The values of the exponents, given by the slopes
of the lines are $\alpha=1.23\pm0.02$ and $\beta=0.81\pm0.02$.}\label{fig5}
\end{figure}
Unlike existing algorithms such as the Configuration Model,
which is affected by uncontrolled biases and unacceptably
long running times except for a very restricted class of
sequences, our algorithm is rejection-free.
Also, it guarantees the independence of the produced samples,
unlike MCMC methods, which have unknown mixing times.
While its mathematical underpinnings
are nontrivial, the algorithm itself is straightforward to implement.
In principle, our approach can be extended to include more complex constraints,
such as a given sequence of motifs frequencies, but we have only concentrated
on degree sequences since they are, arguably, the most fundamental of constraints. The
algorithm can also be used to sample from given in- and out-degree distributions,
not just sequences: given such distributions, one first samples a graphical bds from these,
then one applies our algorithm to generate digraphs. In this case,
however, the sample weights (\ref{weight}) must be modified to reflect the
probability of the occurrence of the given graphical bds when drawn from
the distributions.

Just as in the case of undirected graphs, we can expect the distributions
of the weights for large graphs to be log-normal, as shown in Ref.~\cite{Del10}.
As an example, figure~\ref{fig:logweight} shows the distribution for bi-degree
sequences in which the in-degrees follow a power law with exponent $\gamma=3$
and the out-degrees a Poisson distribution whose mean matches the average
in-degree. Indeed, the distribution of the weight logarithms is well approximated
by a Gaussian. Similarly the undirected case, we find for all the examples 
we studied numerically, that the standard deviation 
$\sigma$ of the distributions
of weight logarithms grows slower than the mean $m$ with the number of nodes
$N$; see figure~\ref{fig5} showing the scaling of $m$ and $\sigma$
for bi-degree sequences in which both in-degrees and out-degrees follow
a power law distribution with exponent $\gamma=3$. Thus, we may expect that
typically, in the $N\rightarrow\infty$
limit, the rescaled weight distribution becomes a delta function, making
the sampling asymptotically uniform.

Bounds on the complexity of the algorithm could easily be obtained by inspecting
the algorithm, showing a maximum complexity on the order of ${\cal O}(N M)$
where $M$ is the total number of edges, $M = \sum_{i=1}^N \bar{d}^{(o)}_i$.

In developing our results, we also provided an efficient way of implementing
the Fulkerson-Ryser test, whose scope of application goes beyond our present
algorithm, as it can be used in any context to determine whether a bi-degree
sequence is graphical.

\ack
HK was supported in part by the US National Science Foundation (NSF) through grant DMR-1005417 and KEB by the NSF grant DMR-0908286.
ZT and HK were supported in part by  the NSF BCS-0826958, 
HDTRA 201473-35045 and by the Army Research Laboratory
under Cooperative Agreement Number W911NF-09-2-0053.  
The views and
conclusions contained in this document are those of the authors and should
not be interpreted as representing the official policies, either expressed
or implied, of the Army Research Laboratory or the U.S. Government.  The
U.S. Government is authorized to reproduce and distribute reprints for
Government purposes notwithstanding any copyright notation here on.

\end{document}